\documentclass[twocolumn,showpacs,preprintnumbers,amsmath,amssymb]{revtex4}
\usepackage[dvips]{graphicx}
\usepackage{epsfig}
\usepackage{soul}
\begin{document}

\title{Quantum discord for two-qubit X-states}

\author{Mazhar Ali\footnote{Electronic address: mazharaliawan@yahoo.com}$^{1,3}$, A. R. P. Rau$^2$ and G. Alber$^1$}
\affiliation{$^1$Institut f\"{u}r Angewandte Physik, Technische Universit\"{a}t Darmstadt, D-64289, Germany \\
$^2$Department of Physics and Astronomy, Louisiana State University, Baton Rouge, Louisiana 70803, USA \\
$^3$Department of Electrical Engineering, COMSATS Institute of Information Technology, Abbottabad 22060, Pakistan}

\begin{abstract}
Quantum discord, a kind of quantum correlation, is defined as the difference between quantum mutual information and classical correlation in a bipartite system. In general, this correlation is different from entanglement, and quantum discord may be nonzero even for certain separable states. Even in the simple case of bipartite quantum systems, this different kind of quantum correlation has interesting and significant applications in quantum information processing. So far, quantum discord has been calculated explicitly only for a rather limited set of two-qubit quantum states and expressions for more general quantum states are not known. In this paper, we derive explicit expressions for quantum discord for a larger class of two-qubit states, namely, a seven-parameter family of so called X-states that have been of interest in a variety of contexts in the field. We also study the relation between quantum discord, classical correlation, and entanglement for a number of two-qubit states to demonstrate that they are independent measures of correlation with no simple relative ordering between them. 

\end{abstract}


\maketitle

\section{Introduction}

For a given bipartite quantum state, it is important to know whether it is entangled, separable, classically
correlated or quantum correlated. Much effort has been invested in subdividing quantum states into separable and
entangled states (see Refs. \cite{Alber-QI2001, Horodecki-RMP-2009} and references therein).
It is well known that entanglement makes possible tasks in quantum information which are impossible without it \cite{NC-QIQC-2000}.
However, entanglement is not the only type of correlation useful for quantum technology.
Recently, it was found that there are some quantum correlations other than entanglement that also offer some advantage,
for example, quantum non-locality without entanglement \cite{Bennett-PRA59-1999, Horodecki-PRA71-2005, Niset-PRA74-2006}.
In addition, it was shown theoretically \cite{Braunstein-PRL83-1999, Meyer-PRL85-2000, Datta-PRL100-2008},
and later experimentally \cite{Lanyon-PRL101-2008}, that some separable states may also speed up certain tasks
over their classical counterparts.
Therefore, it is desirable to investigate, characterize, and quantify quantum correlations more broadly. 

A bipartite quantum state contains both classical and quantum correlation.
These correlations are quantified jointly by their `quantum mutual information', an information-theoretic
measure of the total correlation in a bipartite quantum state \cite{Groisman-PRA72-2005}.
In particular, if $\rho^{AB}$ denotes the density operator of a composite bipartite system $AB$,
and $\rho^A$ ($\rho^B$) the density operator of part $A$ ($B$), respectively, then the quantum mutual information is defined as
\begin{eqnarray}
\mathcal{I} (\rho^{AB}) = S (\rho^A) + S (\rho^B) - S(\rho^{AB})\, , \label{Eq:QMI}
\end{eqnarray}
where $S(\rho) = - \mathrm{tr} \, ( \rho \, \log_2 \rho )$ is the von Neumann entropy. Moreover, it was shown that quantum mutual information is the maximum amount of information that A(lice) can send securely to B(ob) if a composite correlated quantum state is used as the key for a {\it one-time pad cryptographic system} \cite{Schumacher-PRA74-2006}.

Quantum mutual information may be written as a sum of classical correlation $\mathcal{C}(\rho^{AB})$ and quantum correlation $\mathcal{Q} (\rho^{AB})$, that is, $\mathcal{I} (\rho^{AB}) = \mathcal{C} (\rho^{AB}) + \mathcal{Q} (\rho^{AB})$ \cite{Ollivier-PRL88-2001, Vedral-et-al, Luo-PRA77-2008}. This quantum part $\mathcal{Q}$ has been called quantum discord \cite{Ollivier-PRL88-2001}. It is a
different type of quantum correlation than entanglement because separable mixed states (that is, with no entanglement) can have non-zero quantum discord.
 
Quantum discord is not always larger than entanglement \cite{Luo-PRA77-2008, Li-Luo-PRA-2008}. This indicates that discord is not simply a sum of entanglement and some other nonclassical correlation. Even for the simplest case of two entangled qubits, the relation between quantum discord, entanglement, and classical correlation is not yet clear. For pure states and, surprisingly, for a mixture of Bell states, quantum correlation is exactly equal to entanglement whereas classical correlation attains its maximum value $1$. However, for general two-qubit mixed states, the situation is more complicated. Qubit-qubit entanglement has been characterized and quantified completely whereas quantum discord only for particular cases \cite{Ollivier-PRL88-2001, Vedral-et-al, Luo-PRA77-2008,
Li-Luo-PRA-2008, Oppenheim-PRL89-2002, Kaszlikowski-PRL101-2008, Dillenschneider-PRB78-2008, Sarandy-arXiv}. Quantum discord is a measure of nonclassical correlations that may include entanglement but is an independent measure. We will document with simple examples that the amounts of classical correlation, quantum discord and entanglement bear no simple relationship to each other. 
 
In order to quantify quantum discord, Ollivier and Zurek \cite{Ollivier-PRL88-2001} suggested the use of von Neumann type measurements which consist of one-dimensional projectors that sum to the identity operator. Let the projection operators $\{ B_k\}$ describe a von Neumann measurement for subsystem $B$ only, then the conditional density operator $\rho_k$ associated with the measurement result $k$ is 
\begin{eqnarray}
\rho_k = \frac{1}{p_k} (I \otimes B_k) \rho (I \otimes B_k) \,,
\end{eqnarray}
where the probability $p_k$ equals $\mathrm{tr} [(I \otimes B_k) \rho (I \otimes B_k)]$.
The quantum conditional entropy with respect to this measurement is given by \cite{Luo-PRA77-2008} 
\begin{eqnarray}
S (\rho | \{B_k\}) := \sum_k p_k \, S(\rho_k) \, ,
\label{Eq:QCE}
\end{eqnarray}
and the associated
quantum mutual information of this measurement is defined as
\begin{eqnarray}
\mathcal{I} (\rho|\{B_k\}) := S (\rho^A) - S(\rho|\{B_k\}) \, . \label{Eq:QMIM} 
\end{eqnarray}
A measure of the resulting classical correlations is provided \cite{Ollivier-PRL88-2001, Vedral-et-al, Luo-PRA77-2008, Li-Luo-PRA-2008} by 
\begin{eqnarray}
\mathcal{C}(\rho) := \sup_{\{B_k\}} \, \mathcal{I} (\rho|\{B_k\}) \, . \label{Eq:CC} 
\end{eqnarray}

The obstacle to computing quantum discord lies in this complicated maximization procedure for calculating the classical
correlation because the maximization is to be done over all possible von Neumann measurements of $B$. Once $\mathcal{C}$ is in hand, quantum discord is simply obtained by subtracting it from the quantum mutual information,
\begin{eqnarray}
\mathcal{Q}(\rho) := \mathcal{I}(\rho) - \mathcal{C}(\rho) \, .
\end{eqnarray}

For a general two-qubit X-state, the quantification of quantum discord is still missing with only partial results available for subsets of three parameters \cite{Luo-PRA77-2008, Dillenschneider-PRB78-2008, Sarandy-arXiv}. An extension to five real parameters was considered although applications were limited to a smaller subset \cite{Werlang-PRA-2009}. We provide a method to compute the classical correlation and quantum discord for a general two-qubit X-state which depends on seven real parameters. This class includes the maximally entangled Bell states, `Werner' states \cite{Wer-PRA89} which include both separable and nonseparable states, as well as others. We have evaluated an analytical expression which enables us to compute the classical correlation and quantum discord in terms of the density matrix elements of a given X-state. We examine the
relation between classical correlation, quantum discord, and entanglement for various initial states.

This paper is organized as follows. In Section \ref{X-states}, we describe some basic properties of X-states and how to calculate the classical correlation and quantum discord for them. In Section \ref{relation}, we apply this for various examples of X-states, studying the relation between the classical correlation, quantum discord, and entanglement. We conclude our work in Section \ref{conclusion}. An Appendix presents details while extending also the calculation from von Neumann
measurements to more general positive operator valued measurements (POVM) of one subsystem.

\section{Two-qubit X-states}\label{X-states}

In this section, we limit our discussion to initially prepared arbitrary two-qubit X-states. The density matrix of a two-qubit X-state in
the representation spanned by two-qubit product states 
$|1\rangle = |0 \rangle_A
\otimes |0\rangle_B$,
$|2\rangle = |0\rangle_A \otimes |1\rangle_B$,
$|3\rangle =
|1\rangle_A \otimes |0\rangle_B$,
$|4\rangle = |1\rangle_A \otimes |1\rangle_B$ 
is of the general form
\begin{eqnarray}
\rho_X = \left( 
\begin{array}{cccc}
\rho_{11}
& 0 & 0 & \rho_{14} \\ 
0 & \rho_{22} & \rho_{23} & 0 \\ 
0 & \rho_{32} & \rho_{33}
& 0 \\
\rho_{41} & 0 & 0 & \rho_{44}
\end{array}
\right) \,,
\label{Eq:Xstate}
\end{eqnarray}
that is, $\rho_{12} = \rho_{13} = \rho_{24} = \rho_{34} = 0$. This visual appearance resembling the letter $X$ has led them to be called X-states \cite{YuEberly-2007} but, recently, an underlying symmetry structure of these states has been examined \cite{Rau-JPA09}. 

Eq.~(\ref{Eq:Xstate}) describes a quantum state provided the unit trace and positivity conditions $\sum_{i=1}^4 \rho_{ii} = 1$,  $ \rho_{22} \rho_{33} \geq |\rho_{23}|^2$, and $ \rho_{11} \rho_{44} \geq |\rho_{14}|^2$ are fulfilled. X-states are entangled if and only if either $\rho_{22} \rho_{33} < |\rho_{14}|^2$ or $\rho_{11} \rho_{44} < |\rho_{23}|^2$. Both conditions cannot hold simultaneously \cite{STV-PRA98}. Eq.~(\ref{Eq:Xstate}) is a $7$-real parameter state with three real parameters along the main diagonal and two complex (or four real) parameters at off-diagonal positions.

The eigenvalues of the density matrix $\rho_X$ in Eq.~(\ref{Eq:Xstate}) are given by
\begin{eqnarray}
\lambda_0 = \frac{1}{2} \bigg[ \, (\rho_{11} + \rho_{44}) + \sqrt{(\rho_{11} - \rho_{44})^2 + 4 \, |\rho_{14}|^2} \, \bigg] \, , \nonumber \\
\lambda_1 = \frac{1}{2} \bigg[ \, (\rho_{11} + \rho_{44}) - \sqrt{(\rho_{11} - \rho_{44})^2 + 4 \, |\rho_{14}|^2} \, \bigg] \, , \nonumber \\
\lambda_2 = \frac{1}{2} \bigg[ \, (\rho_{22} + \rho_{33}) + \sqrt{(\rho_{22} - \rho_{33})^2 + 4 \, |\rho_{23}|^2} \, \bigg] \, , \nonumber \\
\lambda_3 = \frac{1}{2} \bigg[ \, (\rho_{22} + \rho_{33}) - \sqrt{(\rho_{22} - \rho_{33})^2 + 4 \, |\rho_{23}|^2} \, \bigg] \, . \label{Eq:Evs}
\end{eqnarray}
The quantum mutual information is given as
\begin{eqnarray}
\mathcal{I} (\rho_X) = S(\rho_X^A) + S(\rho_X^B) + \sum_{j=0}^3 \lambda_j \log_2 \lambda_j \,,
\end{eqnarray}
where $\rho_X^A$ and $\rho_X^B$ are the marginal states of $\rho_X$, and 
\begin{eqnarray}
S(\rho_X^A) = - \bigg[ (\rho_{11}+\rho_{22}) \, \log_2 (\rho_{11} + \rho_{22}) + \nonumber \\ (\rho_{33} + \rho_{44}) \, \log_2
(\rho_{33} + \rho_{44}) \, \bigg] \, , \nonumber \\ 
S(\rho_X^B) = - \bigg[(\rho_{11}+\rho_{33}) \, \log_2 (\rho_{11} + \rho_{33}) + \nonumber \\ (\rho_{22} + \rho_{44}) \, \log_2 (\rho_{22} + \rho_{44}) \bigg] \, .
\label{Eq:entropy}
\end{eqnarray}

After computing the quantum mutual information, we need next to compute the classical correlation $\mathcal{C}(\rho_X)$. We consider projective measurements for subsystem $B$ (the projective measurements for subsystem $A$ give the same results if we restrict to either $\rho_{11} = \rho_{44}$, or to $\rho_{22} = \rho_{33}$). We follow the procedure of \cite{Luo-PRA77-2008} except that we are considering a more general class of states than the three-parameter family of that study.

It is known that any von Neumann measurement for subsystem $B$ can be written as Ref. \cite{Luo-PRA77-2008}
\begin{eqnarray}
B_i = V \, \Pi_i \, V^\dagger : \quad i = 0,1 \, , \label{Eq:VNmsur}
\end{eqnarray}
where $\Pi_i = |i\rangle\langle i |$ is the projector for subsystem $B$ along the computational base $|i\rangle$ and 
$V \in SU(2)$ is a unitary operator with unit determinant. After the measurement, the state $\rho_X$ will change to the ensemble $\{ \rho_i , p_i \}$, where
\begin{eqnarray}
\rho_i := \frac{1}{p_i} (I \otimes B_i) \, \rho_X \, (I \otimes B_i) \,,
\label{Eq:Tauk}
\end{eqnarray}
and $p_i = \mathrm{tr} \, [\, (I \otimes B_i) \, \rho_X \, (I \otimes B_i)\, ]$. The $\{ \rho_i , p_i \}$, with $i=0,1$ are of subsystem A and thus $2 \times 2$ density matrices.  

We may write any $V \in SU(2)$ as
\begin{eqnarray}
V = t \, I + \mathrm{i} \, \vec{y}\cdot\vec{\sigma} \, ,\label{Eq:defV}
\end{eqnarray}
with $t,y_1,y_2,y_3 \in \mathbb{R}$ and $t^2 + y_1^2 + y_2^2 + y_3^2 = 1$. 
This implies that these parameters, three among them independent, assume their values in the interval $[-1,1]$, i.\,e. $t\, , \, y_i \in [-1,1]$ for $i = 1,2,3$.
The ensemble $\{\rho_i , p_i\}$ can be characterized by their eigenvalues as per a derivation given in the Appendix. The two eigenvalues each of $\rho_0$ and $\rho_1$ are given as
\begin{eqnarray}
v_\pm (\rho_0) = \frac{1}{2} (1 \pm \theta) \,, \nonumber \\
w_\pm (\rho_1)= \frac{1}{2} (1 \pm \theta') \, . \label{Eq:eval}
\end{eqnarray}
The corresponding probabilities are given as
\begin{eqnarray}
p_0 &=& [\, (\rho_{11} + \rho_{33}) \, k + (\rho_{22} + \rho_{44}) \, l \, ] \,, \nonumber \\
p_1 &=& [ \, (\rho_{11} + \rho_{33}) \, l + (\rho_{22} + \rho_{44}) \, k \, ] \,. \label{Eq:prob}
\end{eqnarray}
We have defined $\theta$ and $\theta'$ that generalize a single expression in Ref. \cite{Luo-PRA77-2008}  as 
\begin{eqnarray}
\theta = \sqrt{ \frac{[\, (\rho_{11}-\rho_{33}) \, k + (\rho_{22} - \rho_{44}) \, l \, ]^2 + \Theta }{[\, (\rho_{11}+\rho_{33}) \, k + (\rho_{22} + \rho_{44}) \, l \, ]^2}} \, , \label{Eq:theta} \\
\theta' = \sqrt{ \frac{[\, (\rho_{11}-\rho_{33}) \, l + (\rho_{22} - \rho_{44}) \, k \, ]^2 + \Theta }{[\, (\rho_{11}+\rho_{33}) \, l + (\rho_{22} + \rho_{44}) \, k \, ]^2}} \,,
\label{Eq:thetap}
\end{eqnarray}
where $\Theta = 4 \, k \, l \, [\, |\rho_{14}|^2 + |\rho_{23}|^2 + 2 \, \Re(\rho_{14} \rho_{23})\, ] - 16 \, m \, \Re(\rho_{14} \rho_{23}) + 16 \, n \,  \Im(\rho_{14} \rho_{23}) $, and $\Re(z)$ and $\Im(z)$ are the real and imaginary parts of the complex number $z$. 
We have defined the parameters, $m$, $n$, $k$, and $l$ as
\begin{eqnarray}
m =& (t \, y_1 + y_2 \, y_3)^2 \,, \, \, n = (t \, y_2 - y_1 \, y_3)(t \, y_1 + y_2 \, y_3)\,, \nonumber \\& k = t^2 + y_3^2 \,, \quad l = y_1^2 + y_2^2 \, . \label{Eq:kmn}
\end{eqnarray}
With $k + l = 1$, Eqs.~(\ref{Eq:theta}) and (\ref{Eq:thetap}) for a given density matrix depend on three real parameters $k$, $m$, and $n$. It can be readily checked that $k \in [0,1]$, $m \in [0,1/4]$, and $n \in [-1/8,1/8]$. These three parameters are a recasting of the three independent parameters in Eq.~(\ref{Eq:defV}) and are related to the set $(z_1, z_2, z_3)$ of Ref. \cite{Luo-PRA77-2008} through $4m=z_2^2, 4n = -z_1z_2, k-l =z_3$.  

The entropies of the ensemble $\{\rho_i , p_i\}$ are given as
\begin{eqnarray}
S (\rho_0) = - \frac{1-\theta}{2} \log_2 \frac{1-\theta}{2} - \frac{1+\theta}{2} \log_2 \frac{1+\theta}{2} \, , \label{Eq:Entemb0}
\end{eqnarray}
\begin{eqnarray}
S (\rho_1) = - \frac{1-\theta'}{2} \log_2 \frac{1-\theta'}{2} - \frac{1+\theta'}{2} \log_2 \frac{1+\theta'}{2} \, .
\label{Eq:Entemb1}
\end{eqnarray}
The quantum conditional entropy in Eq.~(\ref{Eq:QCE}) is given as
\begin{eqnarray}
S (\rho_X|\{B_i\}) = p_0 \, S(\rho_0) + p_1 \, S(\rho_1) \, . \label{Eq:qcetx}
\end{eqnarray}
As per Eq.~(\ref{Eq:CC}), the classical correlation is obtained as
\begin{eqnarray}
\mathcal{C}(\rho_X) &=& \sup_{\{B_i\}} \, [ \, \mathcal{I}(\rho_X|\{B_i\}) \, ] \nonumber \\ &=& S(\rho_X^A) - \min_{\{B_i\}} \, [
\, S (\rho_X|\{B_i\})\, ] \, . \label{Eq:qmimtaux}
\end{eqnarray}
Therefore, to calculate the classical correlation and consequently quantum discord, we have to minimize the quantity $S (\rho_X|\{B_i\})$ (Eq.~(\ref{Eq:qcetx})) with respect to the von Neumann measurements. 

To minimize Eq.~(\ref{Eq:qcetx}) by setting equal to zero its partial derivatives with respect to $k, m$ and $n$, observe first that the expression is symmetric under the interchange of $k$ and $l=1-k$. It is, therefore, an even function of $(k-l)$ and the extremum lies at $k=l=1/2$ or at the end points $k=0$ or $k=1$. From the definition of these parameters in Eq.~(\ref{Eq:kmn}), the end points require $t=y_3=0$ or $y_1=y_2=0$ and, therefore, $m=n=0$. On the other hand, for the case $k=l=1/2$, we have $\theta = \theta'$, and $S(\rho_0) = S(\rho_1)$ and the minimization of $S (\rho_X|\{B_i\})$ is equal to the minimization of either $S(\rho_0)$ or $S(\rho_1)$. Further, with $\Theta$ in these expressions involving $m$ and $n$ only linearly, extreme values are attained only at their end points: $m = 0,\, 1/4$ and $n=0, \, \pm 1/8$. Thus, the maximum classical correlation and thereby the quantum discord can be obtained easily analytically. The Appendix shows how we may generalize to positive operator valued measurements (POVM) to get final compact expressions that are simple extensions of the more limited von Neumann measurements, thereby yielding the same value for the maximum classical correlation and discord.

For the special case of two-qubit X-states with restrictions $\rho_{11} = \rho_{44}$, $\rho_{22} = \rho_{33}$, and with real off-diagonal elements, we define 
\begin{eqnarray}
\theta_1 = 2 \, | \rho_{14} + \rho_{23}| \, , \quad \theta_2 = 2 \, |\rho_{14} - \rho_{23}| \, , \nonumber \\ \theta_3 =
\theta_4 = |(\rho_{11} + \rho_{44})-(\rho_{22} + \rho_{33})|, 
\end{eqnarray}
and associated entropy $S'(\rho_i)|_{\theta_j}$ as
\begin{eqnarray}
S'(\rho_i)|_{\theta_j} = - \frac{1 + \theta_j}{2} \, \log_2 \frac{1 + \theta_j}{2} - \frac{1 - \theta_j}{2} \,
\log_2 \frac{1 - \theta_j}{2} \,. \nonumber \\ \label{Eq:min2}
\end{eqnarray}
In addition, $S(\rho_0) = S(\rho_1)$, and the minimum value of $S(\rho|\{B_i\})$ is equal to the minimum value of $S(\rho_0)$, which is given as
\begin{eqnarray}
\min [ \, S(\rho|\{B_i\}) \, ] = \min [\, S(\rho_0) \, ] = S'(\rho_0)|_{\theta_{sup}} \, ,
\end{eqnarray}
where 
\begin{eqnarray}
\theta_{sup} = \max \, \{ \, \theta_1 , \theta_2 , \theta_3 \, \} \ .
\end{eqnarray}
Therefore, we recover the results of Ref. \cite{Luo-PRA77-2008} as a special case of ours.

\section{Relation between discord and entanglement}\label{relation}

In this section, we study the relation between the classical correlation, quantum discord, and entanglement for various initial states. 

$(1)$ As a first example, we take maximally entangled pure states, that is, the four Bell states given as $|\psi^\pm\rangle = (|0,1\rangle \pm |1,0\rangle)/\sqrt{2}$, and $|\phi^\pm\rangle = (|0,0\rangle \pm |1,1\rangle)/\sqrt{2}$. It is well known that for any Bell state, we
have
\begin{eqnarray}
\mathcal{I}(\rho) = 2\,,\quad \mathcal{C}(\rho) = 1\,, \quad \mathcal{Q}(\rho) = 1 \, . 
\end{eqnarray}
For this particular case, quantum discord and any measure of entanglement coincide and are equal to the maximum value of the correlation. Surprisingly, if we mix any two Bell states, for example, $\rho = a |\psi^+\rangle\langle\psi^+| + (1-a) |\phi^+\rangle\langle\phi^+|$, the classical correlation is not affected at all, and quantum discord is captured by a measure of entanglement called the entanglement of formation \cite{Bennett-PRA54-1996, Wootters-PRL80-1998}. The quantum mutual information is $\mathcal{I}(\rho) = 2 + a \, \log_2 a + (1-a) \log_2(1-a)$, the classical correlation is $\mathcal{C}(\rho) = 1$, entanglement of formation is equal to quantum discord, i.\,e., $\mathcal{E}(\rho) = \mathcal{Q}(\rho) = 1 + a \, \log_2 a + (1-a) \log_2(1-a)$ \cite{Vedral-et-al}. However, this example is special as for other mixed states, entanglement and quantum discord differ substantially.

$(2)$ We take the class of states defined as $\rho = a \, |\psi^+\rangle\langle\psi^+| + (1- a) \, |1,1\rangle\langle1,1|$ ($0 \leq a \leq 1$), where $|\psi^+\rangle = (|0,1\rangle + |1,0\rangle)/\sqrt{2}$, is a maximally entangled state. Based on the results of the previous section, we are now able to calculate the classical correlation and quantum discord. We have $\theta_1 = \theta_2 = \sqrt{a^2 + (1-a)^2}$, $\theta_3 = |2 -3 \, a |/(2-a)$, and $\theta_4 = 1$. As $\theta_3 \neq \theta_4$, the quantity $S(\rho|\{B_i\})|_{\theta_3,\theta_4}$ is given as
\begin{eqnarray}
S(\rho|\{B_i\})|_{\theta_3,\theta_4} =  \frac{2-a}{2} \, S'(\rho_0)|_{\theta_3} + \frac{a}{2} \, S'(\rho_1)|_{\theta_4} \, .
\end{eqnarray}
As $\theta_4 = 1$, we have $S'(\rho_1)|_{\theta_4} = 0$. We note that $S'(\rho_0)|_{\theta_1} \leq S(\rho|\{B_i\})|_{\theta_3,\theta_4}$, i.\,e., $\min \, \{ \, S'(\rho_0)|_{\theta_1} \,, \, S(\rho|\{B_i\})|_{\theta_3,\theta_4} \, \} = S'(\rho_0)|_{\theta_1} $. 

The classical correlation is given as
\begin{eqnarray}
\mathcal{C}(\rho) = S(\rho^A) - S'(\rho_0)|_{\theta_1}\,,
\end{eqnarray}
where
\begin{eqnarray}
S(\rho^A) = -\frac{a}{2} \, \log_2 \frac{a}{2} - \frac{2-a}{2} \, \log_2 \frac{2-a}{2} \, .
\end{eqnarray}
The quantum mutual information is 
\begin{eqnarray}
\mathcal{I}(\rho) = 2 \, S (\rho^A) - S(\rho) \,, 
\end{eqnarray}
and quantum discord
\begin{eqnarray}
\mathcal{Q}(\rho) = S(\rho^A) + S'(\rho_0)|_{\theta_1} - S(\rho) \, , 
\end{eqnarray}
where 
\begin{eqnarray}
S(\rho) = - a \, \log_2 a - (1-a) \, \log_2 (1-a) \, . 
\end{eqnarray}

To study the relation between quantum discord and entanglement, we choose a measure of entanglement. Although various measures of entanglement \cite{Bennett-PRA54-1996, Wootters-PRL80-1998,Vidal-PRA65-2002} give the same result for separable states and for Bell states, the amount of entanglement of a specific mixed state is different for different measures. We prefer to compare quantum discord with concurrence $C'$ \cite{Wootters-PRL80-1998}. The concurrence for this state is given as
\begin{eqnarray}
C'(\rho) =  a \, . \label{Eq:con1}  
\end{eqnarray}

\begin{figure}[h]
\scalebox{2.0}{\includegraphics[width=1.7in]{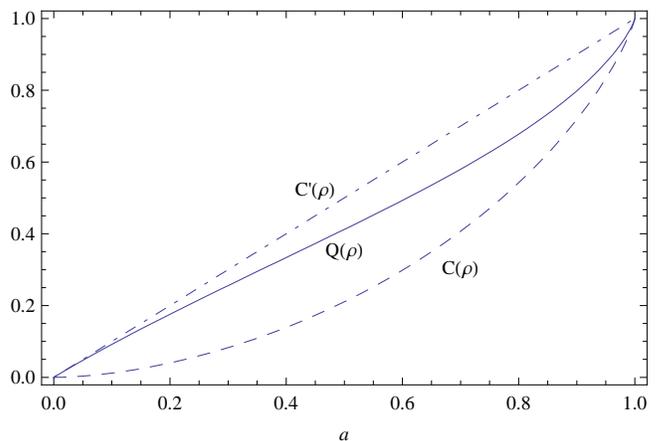}}
\caption{Concurrence, quantum discord, and classical correlation for the class of states in $(2)$ of the adjoining text are plotted for $\rho$ as a function of the parameter $a$. These correlations are equal only for $a = 0$ and $a = 1$.}
\label{Fig:1}
\end{figure}

Figure \ref{Fig:1} displays the classical correlation, quantum discord, and concurrence for $\rho$ for various values of the parameter $a$. The solid line presents quantum discord, the dotted-dashed line is for concurrence, whereas the dashed line is for classical correlation. It can be seen that for this particular initial state, quantum discord is always less than concurrence but always greater than the classical correlation.

$(3)$ We consider the initial state $\rho = a \, |\phi^+\rangle \langle \phi^+| + (1-a) \, |1,1\rangle\langle 1,1|$ ($0 < a \leq 1$), where $|\phi^+\rangle = (|0,0\rangle + |1,1\rangle)/\sqrt{2}$, is a maximally entangled state. In this case $\theta_1 = \theta_2 =\sqrt{a^2 + (a-1)^2 }$, and $\theta_3 = \theta_4 = 1$. Therefore, $S(\rho|\{B_i\})|_{\theta_3,\theta_4} = 0$, which is the minimum when compared with $S'(\rho_0)|_{\theta_1}$, that is, $ \min \, \{ \, S'(\rho_0)|_{\theta_1} \,, \, S(\rho|\{B_i\})|_{\theta_3,\theta_4} \, \} = 0$. Hence the classical correlation is given as
\begin{eqnarray}
\mathcal{C}(\rho) = S(\rho^A) = - \frac{a}{2} \, \log_2 \frac{a}{2} - \frac{2-a}{2} \log_2 \, \frac{2-a}{2} \,.
\end{eqnarray}
The quantum mutual information is given as
\begin{eqnarray}
\mathcal{I}(\rho) = 2 \, S(\rho^A) - S (\rho) \, ,
\end{eqnarray}
and quantum discord as 
\begin{eqnarray}
\mathcal{Q}(\rho) = S(\rho^A) - S(\rho) \,, 
\end{eqnarray}
with  
$S(\rho)$ as in Eq.~(\ref{Eq:min2}) . The concurrence for these states is again given as in Eq.~(\ref{Eq:con1}),  
$C'(\rho) = a$.

\begin{figure}[h]
\scalebox{2.0}{\includegraphics[width=1.7in]{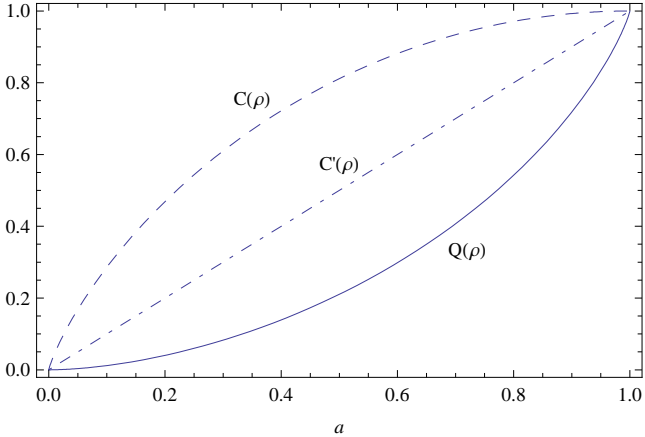}}
\caption{As in Fig. 1 for the class of states in $(3)$ in the text. Top curve (dashed line) is for classical correlation, middle one (dotted-dashed) is for concurrence, and bottom curve (solid line) is for quantum discord.}
\label{Fig:2}
\end{figure}

The classical correlation, quantum discord, and concurrence have been plotted in Figure \ref{Fig:2} for $\rho$ against parameter $a$. The solid line is for quantum discord, the dotted-dashed line is for concurrence, and the dashed line for classical correlation. Interestingly, for this particular initial state, the classical correlation is always greater than both entanglement and quantum discord except for $a = 0$ and $a = 1$. Moreover, quantum discord is always less than concurrence and the classical correlation.

$(4)$ Let us consider the Werner state \cite{Wer-PRA89}
\begin{eqnarray}
\rho = a \, |\psi^-\rangle\langle\psi^-| + \frac{1-a}{4} \, I \,,
\label{Eq:W-State}
\end{eqnarray}
where $|\psi^-\rangle = (|0,1\rangle - |1,0\rangle)/\sqrt{2}$ is a maximally entangled state and $ 0 \leq a \leq 1$. The Werner state has a peculiar property that the eigenvalues of its ensemble $\{ \, p_i \,, \rho_i \, \}$ do not depend on the parameters $k$ and $m$, i.\,e. $\theta = \theta' = a$ or in other words $\theta_1 = \theta_2 = \theta_3 = a$, and we have \cite{Luo-PRA77-2008}
\begin{eqnarray}
\mathcal{C}(\rho) = \frac{1-a}{2} \, \log_2 (1-a) + \frac{1+a}{2} \, \log_2 (1+a) \,,
\end{eqnarray}
\begin{eqnarray}
\mathcal{I}(\rho) =  \frac{3(1-a)}{4} \log_2 (1-a) + \frac{1 + 3 a}{4} \log_2 (1 + 3 a) \,,
\end{eqnarray}
and quantum discord
\begin{eqnarray}
\mathcal{Q}(\rho) &=& \mathcal{I}(\rho) - \mathcal{C}(\rho) \nonumber \\ & = & \frac{1}{4} \big[ \, (1-a) \log_2 (1-a) + (1 + 3 a) \log_2 (1 + 3 a) \nonumber \\&&  - 2 \,(1+a) \, \log_2 (1+a) \, \big] \, .
\end{eqnarray}
The concurrence for the Werner state is given by
\begin{eqnarray}
C'(\rho) = \max \, \big\{ \, 0 \,, \frac{3 \, a - 1}{2} \, \big\} \, . 
\end{eqnarray}

\begin{figure}[h]
\scalebox{2.0}{\includegraphics[width=1.7in]{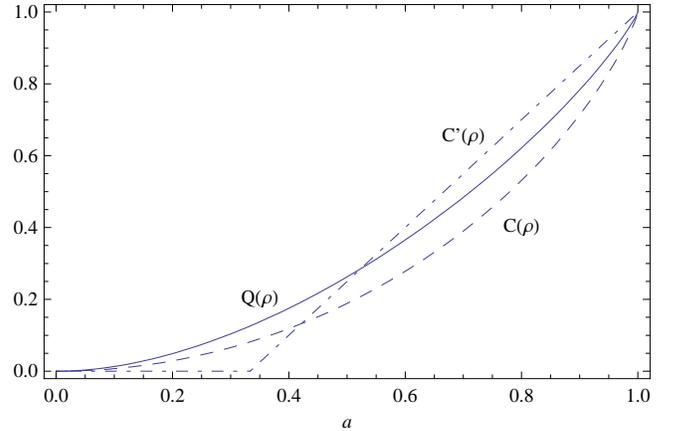}}
\caption{Graphs of quantum discord $\mathcal{Q}(\rho)$ (solid line), classical correlation $\mathcal{C}(\rho)$ (dashed line), and concurrence $C'(\rho)$ (dotted-dashed line) versus $a$ for Werner state.}
\label{Fig:3}
\end{figure}

These correlations are plotted in Figure \ref{Fig:3}. In contrast to previous examples, the correlations have a different order as functions of $a$, with quantum discord initially larger than concurrence and the classical correlation for $0 \leq a \lesssim 0.523$, but for the range $0.523 < a < 1$, concurrence becoming larger than discord and the classical correlation. Such a behavior is different from the
previous two examples.

$(5)$ Finally we consider the initial state 
\begin{eqnarray}
\rho = \frac{1}{3} \, \{ \, (1-a) |0,0\rangle\langle0,0| + 2 \, |\psi^+\rangle\langle \psi^+ | + a |1,1\rangle\langle1,1| \, \} \, .\nonumber \\
\end{eqnarray}
In contrast with example ($2$), this state also contains the $|0,0\rangle$ noisy component. We note that $\theta_1 = \theta_2 =  \sqrt{(1- 2 a)^2 + 4}/3$, $\theta_3 = (1-a)/(1+a)$, and $\theta_4 = a/(2-a)$. We find $S'(\rho_0)|_{\theta_1} < S(\rho|\{B_i\})|_{\theta_3,\theta_4}$, that is, $\min \, \{ \, S'(\rho_0)|_{\theta_1} \, , \, S(\rho|\{B_i\})|_{\theta_3,\theta_4} \, \} =
S'(\rho_0)|_{\theta_1} $. We have 
\begin{eqnarray}
\mathcal{C}(\rho) = \frac{1 + \theta_1}{2} \, \log_2 \frac{1 + \theta_1}{2} + \frac{1 - \theta_1}{2} \, \log_2 \frac{1 - \theta_1}{2} \nonumber \\ - \frac{2 - a}{3} \, \log_2 \frac{2 - a}{3} - \frac{1 + a}{3} \, \log_2 \frac{1 + a}{3} \, ,
\end{eqnarray}
\begin{eqnarray}
\mathcal{I}(\rho) = \frac{1 - a}{3} \, \log_2 \frac{1 - a}{3} + \frac{a}{3} \, \log_2 \frac{a}{3} + \frac{2}{3} \, \log_2 \frac{2}{3} - \nonumber \\ \frac{2(2 - a)}{3} \, \log_2 \frac{2 - a}{3} - \frac{2 (1 + a)}{3} \, \log_2 \frac{1 + a}{3} \,,
\end{eqnarray}
and quantum discord
\begin{eqnarray}
\mathcal{Q}(\rho) = \frac{1 - a}{3} \, \log_2 \frac{1 - a}{3} + \frac{a}{3} \, \log_2 \frac{a}{3} + \frac{2}{3} \, \log_2 \frac{2}{3}
\nonumber \\ - \frac{(2 - a)}{3} \, \log_2 \frac{2 - a}{3} - \frac{ (1 + a)}{3} \, \log_2 \frac{1 + a}{3} \nonumber \\ - \frac{1 + \theta_1}{2} \, \log_2 \frac{1 + \theta_1}{2} - \frac{1 - \theta_1}{2} \, \log_2 \frac{1 - \theta_1}{2} \, .
\end{eqnarray}
The concurrence for this state is given as
\begin{eqnarray}
C'(\rho) = \max \, \big\{ \, 0 \, , \frac{2}{3} \big[ \, 1 - \sqrt{a (1-a)} \, \big] \, \big\} \, . 
\end{eqnarray}

\begin{figure}[h]
\scalebox{2.0}{\includegraphics[width=1.7in]{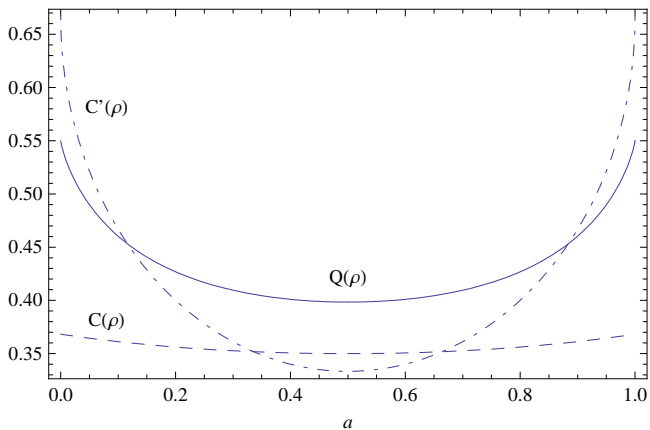}}
\caption{Graphs of quantum discord $\mathcal{Q}(\rho)$ (solid line), classical correlation $\mathcal{C}(\rho)$ (dashed line), and concurrence $C'(\rho)$ (dotted-dashed line) versus $a$. The correlations are  symmetrical about $a = 0.5$ and are maximum at $a = 0$ and $a = 1$.}
\label{Fig:4}
\end{figure}

We display the classical correlation, quantum discord, and concurrence versus $a$ for this state in Figure \ref{Fig:4}. We can see that all these correlations are symmetric about $a = 1/2$. As the parameter $a$ varies from $0$ to $1$, the quantum mutual information decreases, and the classical correlation, quantum discord, and
entanglement decrease, and vice versa.

\section{Summary}\label{conclusion}

We have derived analytical expressions for the classical correlation and quantum discord in X-states, a seven-parameter family of states of two qubits. This generalizes results previously available only for a three-parameter subset of such states. A large class of two-qubit states that includes maximally or partially entangled states, and mixed states that are separable or non-separable can now be examined for these various correlations. We present such results for the classical correlation, quantum discord, and entanglement for various density matrices. We conclude that there are no simple ordering relations between these correlations and, in particular, that quantum discord may be smaller or larger than entanglement as measured with concurrence or negativity. Thus, quantum discord is a fundamentally different resource than entanglement and these correlations are different qualitatively and quantitatively. 

\acknowledgments{
We thank Drs. Jaroslav Novotn\'y and Joseph Rennes for discussions. MA would like to thank Prof. Christof Wunderlich for his kind hospitality at Universit\"at Siegen where part of this work is done. MA acknowledges financial support by the Higher Education Commission, Pakistan, and by the DAAD. ARPR acknowledges support from the Alexander von Humboldt Foundation.
}

\section*{Appendix: Calculational details of von Neumann and POVM alternatives}

In this appendix, we give some details of the calculation leading to Eqs.~(\ref{Eq:prob}) - (\ref{Eq:Entemb1}). First, an alternative 7-parameter version of the density matrix in Eq.~(\ref{Eq:Xstate}) proves useful:
\begin{eqnarray}
\rho_X = \frac{1}{4} \left( 
\begin{array}{cccc}
1+d_1
& 0 & 0 & c_1 -c_2 \\ 
0 & 1+d_2 & c_1+c_2 & 0 \\ 
0 & c_1^* +c_2^* & 1+d_3
& 0 \\
c_1^* -c_2^* & 0 & 0 & 1+d_4
\end{array}
\right) \,,
\label{Eq:AXstate}
\end{eqnarray}
where $c_1$ and $c_2$ are complex and the other coefficients real, the diagonal entries being $(d_1 = c_3 + a_3 + b+3$, $d_2 = -c_3 + a_3 - b_3$, $d_3 = -c_3 - a_3 + b_3$, $d_4 = c_3 - a_3 - b_3)$. These parameters are given in terms of the density matrix elements by $c_3 = \rho_{11} + \rho_{44} - \rho_{22} - \rho_{33}$, $a_3 = \rho_{11} - \rho_{44} + \rho_{22} - \rho_{33}$, $b_3 = \rho_{11} - \rho_{44} - \rho_{22} + \rho_{33}$, $c_1 = 2 (\rho_{23} + \rho_{14})$, and $c_2 = 2 (\rho_{23} - \rho_{14})$. The choice of only three coefficients $c_i$, all real, corresponds to the one made in Ref. \cite{Luo-PRA77-2008}.    

Evaluation of Eq.~(\ref{Eq:Tauk}) as per the procedure in Ref. \cite{Luo-PRA77-2008} gives after some algebra,
\begin{eqnarray}
p_0 \rho_0 \!\!&=& \!\frac{1}{4} [1+b_3z_3+a_1\sigma_1+a_2\sigma_2+(a_3\!+\!c_3z_3)\sigma_3] , \nonumber \\
p_1\rho_1 \!\! &=& \! \frac{1}{4} [1-b_3z_3-a_1\sigma_1-a_2\sigma_2+(a_3\!-\!c_3z_3)\sigma_3] . \label{Eq:Aprob}
\end{eqnarray}
with 
\begin{eqnarray}
a_1 &=& \Re(z_1c_1-iz_2c_2) = z_1 \, \Re(c_1) + z_2 \, \Im(c_2), \nonumber \\
a_2 &=& \Im(z_1c_1-iz_2c_2) = z_2 \, \Re(c_2) - z_1 \, \Im(c_1), \label{Eq:defa}
\end{eqnarray}
in terms of the unit vector $\vec{z}$ defined in Ref. \cite{Luo-PRA77-2008} from the parameters in Eq.~(\ref{Eq:defV}) as
\begin{eqnarray}
z_1 =& 2 \, (-t \, y_2 + y_1 \, y_3) \,, \quad z_2 = 2 \, (t \, y_1 + y_2 \, y_3) \, , \nonumber \\& z_3 = t^2 + y_3^2 - y_1^2 - y_2^2. 
\label{Eq:z123}
\end{eqnarray}
Upon taking the trace of Eq.~(\ref{Eq:Aprob}), we get the probabilities
\begin{equation}
p_0 = (1 + b_3 \, z_3)/2 \,, \quad p_1 = (1 - b_3 \, z_3)/2 \,, \label{Eq:Aprob2}
\end{equation}
in agreement with Eq.~(\ref{Eq:prob}) upon writing $b_3$ in terms of $\rho_{ij}$ and $z_3=k-l$. 

The density matrices themselves for subsystem A that follow from Eq.~(\ref{Eq:Aprob}) are, therefore,

\begin{eqnarray}
\rho_0 \!\!&=& \!\!\frac{1}{2} \left( I \!+ \![a_1 \sigma_1 +a_2 \sigma_2 +(a_3\!+c_3z_3)\sigma_3]/(1\!+b_3z_3) \right), \nonumber \\
\rho_1\!\!&=& \!\!\frac{1}{2} \left( I \!+ \![-a_1 \sigma_1 -a_2 \sigma_2 +(a_3\!-c_3z_3)\sigma_3]/(1\!-b_3z_3) \right).\nonumber\\
&&
\label{Eq:Arho}
\end{eqnarray}
Their eigenvalues are the ones in Eq.~(\ref{Eq:eval}). Note that the denominators in Eq.~(\ref{Eq:thetap}) are $(1 \pm b_3 \, z_3)/2$, while the numerators under the square root are the sum of the squares of $(a_3 \pm c_3 \, z_3)/2$ and $\Theta = (a_1^2 + a_2^2)/4$ as follows from the properties of the Pauli matrices. The expressions for $\rho_0$ and $\rho_1$ differ in a change in sign in $\vec{z}$ in Eq.~(\ref{Eq:Arho}) and Eq.~(\ref{Eq:defa}). This close connection except for a change in sign of $\vec{z}$ traces back to the similar change in sign of the two von Neumann projectors in Eq.~(\ref{Eq:VNmsur}), $\Pi_{0,1} = (I \pm \sigma_z)/2$. Indeed, the reduction of the lengthy calculation to this compact statement of passing from the projectors $\Pi_{0,1}$ to the expressions in Eq.~(\ref{Eq:Arho}), with the measurement directions
$\pm \hat{z}$ replaced by the dot product of $\vec{\sigma}$ with $\vec{z}$ and corresponding coefficients of the density matrix, will be useful in generalizing to other types of measurements below.

Instead of von Neumann projectors, consider more general positive operator valued measurements (POVM). For instance, choose three orthogonal unit vectors mutually at 120$^o$,

\begin{equation}
\hat{s}_{0,1,2} = [ \hat{z} \, , (- \hat{z} \pm \sqrt{3} \, \hat{x})/2],
\label{Eq:vecs}
\end{equation}
and corresponding projectors
\begin{eqnarray}
E_0 &=& \frac{1}{3} (I + \sigma_z), \nonumber \\
E_{1,2} &=& \frac{1}{3}(I - \frac{1}{2} \sigma_z \pm \frac{\sqrt{3}}{2} \sigma_x),
\label{Eq:POVMProj}
\end{eqnarray}
with $E_0 + E_1 + E_2 = I$, $E_i^2 = 2 \, E_i/3$, ${\rm tr}(E_i) = 2/3$.
With this choice of projectors in Eq.~(\ref{Eq:VNmsur}), the same calculation for what remains for subsystem A after the measurements on subsystem B give now the counterparts of Eq.~(\ref{Eq:Aprob2}),
\begin{eqnarray}
p_0 =& (1 + b_3 \, z_3)/3 \, , \quad p_1 = (1 + b_3 \, \alpha_3)/2 \, , \nonumber \\& 
p_2 = (1 + b_3 \, \beta_3)/3, \label{Eq:Aprob3}
\end{eqnarray}
and of Eq.~(\ref{Eq:Arho}) for $\rho_{0.1.2}$ with exactly similar expressions except $(\vec{z}, \vec{\alpha}, \vec{\beta})$ replace $\pm \vec{z}$ in Eq.~(\ref{Eq:Arho}) and Eq.~(\ref{Eq:defa}). Here, we have defined two counterparts of $\vec{z}$ that follow from the other two projectors in Eq.~(\ref{Eq:vecs}),
\begin{equation}
\vec{\alpha} \, , \, \vec{\beta} = (- \vec{z} \pm \sqrt{3} \, \vec{x})/2 \, ,
\end{equation}
with $\vec{z}$ given in Eq.~(\ref{Eq:z123}) in terms of the parameters in Eq.~(\ref{Eq:defV}) and similarly we have defined $\vec{x}$ drawn from another set of combinations in Ref. \cite{Luo-PRA77-2008},
\begin{eqnarray}
x_1 =& t^2 + y_1^2 - y_2^2 - y_3^2 \, , \quad x_2 = 2 \, (-t \, y_3 + y_1 \, y_2) \, , \nonumber \\& 
x_3 = 2 \, (t \, y_2 + y_1 \, y_3) \, . 
\label{Eq:x123}
\end{eqnarray}
Other choices of projectors may define a similar third set of parameters $\vec{y}$ in Ref. \cite{Luo-PRA77-2008}, all these unit vectors $(\vec{x}, \vec{y}, \vec{z})$ being mutually orthogonal.

\end{document}